\documentclass[preprintnumbers, floatfix, twocolumn,
preprintnumbers, letterpaper, superscriptaddress,nofootinbib]{revtex4}
\usepackage{eurosym}
\usepackage{amsfonts}
\usepackage{amsmath}
\usepackage{amssymb,epsf}
\usepackage{latexsym}
\usepackage{graphicx,epsfig}
\usepackage{amssymb}
\usepackage{subfigure}
\usepackage[colorlinks=true,citecolor=blue,linkcolor=blue,urlcolor=black]{hyperref}

\graphicspath{{Images/}}
\begin{document}

\title{Charged scalar quasi-normal modes for \\
higher-dimensional Born-Infeld dilatonic black holes with Lifshitz scaling}
\author{S. Sedigheh Hashemi}
\affiliation{School of Particles and Accelerators, Institute for Research in Fundamental Sciences (IPM), P. O. Box:
19395-5531, Tehran, Iran}
\author{Mahdi Kord Zangeneh}
\email{mkzangeneh@scu.ac.ir}
\affiliation{Physics Department, Faculty of Science, Shahid Chamran University of Ahvaz,
Ahvaz 61357-43135, Iran}
\affiliation{Research Institute for Astronomy and Astrophysics of Maragha (RIAAM), P. O.
Box: 55134-441, Maragha, Iran}
\author{Mir Faizal}
\email{mirfaizalmir@googlemail.com}
\affiliation{Department of Physics and Astronomy, University of Lethbridge, Lethbridge,
AB T1K 3M4, Canada}
\affiliation{Irving K. Barber School of Arts and Sciences, University of British Columbia
- Okanagan, Kelowna, BC, V1V 1V7, Canada}

\begin{abstract}
We study quasi-normal modes for a higher dimensional
black hole with Lifshitz scaling, as these quasi-normal 
modes can be used to test Lifshitz models with large
extra dimensions. Furthermore, as the effective Planck scale
is brought down in many models with large extra dimensions, we study these
quasi-normal modes for a UV completion action. Thus, we analyze quasi-normal
modes for higher dimensional dilaton-Lifshitz black hole solutions coupled
to a non-linear Born-Infeld action. We will analyze the charged
perturbations for such a black hole solution. We will first analyze the
general conditions for stability analytically, for a positive potential.
Then, we analyze this system for a charged perturbation as well as negative
potential, using the asymptotic iteration method for quasi-normal modes.
Thus, we analyze the behavior of these modes numerically.
\end{abstract}

\maketitle

\section{Introduction}

Mathematically ideal black holes are usually studied as isolated systems.
However, this mathematical idealization cannot be actualized in any real
world system. This is because black holes will always interact with matter
surrounding it. Even if all the matter around a black hole could be removed,
it would still interact with the vacuum around it, and thus cannot be
considered an isolated system. So, in any realistic description of a black
hole we would have to consider such perturbative effects on the basic
parameters describing such a black hole. Such perturbations of a black hole
lead to the emission of gravitational waves \cite{1102.4014}. Initially
there is an short outburst, which is followed by long period of damping
oscillations in the form of quasi-normal modes (QNMs) \cite%
{9909058,159,0511133}, and finally these QNMs are suppressed at late times
by power-law or exponential tails. A complex frequency called the
quasi-normal frequency (QNF) is associated with these QNMs \cite%
{Regge,FJ,FJ2}. The real part of this complex frequency represents the
frequency of oscillation, and its imaginary part represents the rate of
damping of that oscillation.

In this paper, we will be analyzing the QNMs as they are the most
interesting phase of radiation, which can have interesting physical
consequences. This is because the study of gravitational wave has become
important due to the detection of these wave in GW150914 \cite%
{Abbott:2016blz}, and the analysis of QNMs from the data obtained from
gravitational antennas such as LIGO, VIRGO and LISA, can have important
consequences \cite{z2, z3, z4, z5}. The existence of extra dimensions in
string theory has motivated the study of black hole in large
extra-dimensions, such that the the extra dimensions are large enough to
suppress the Planck scale to the $TeV$ scale \cite{m1, m2, 0106219,
0806.3381}. These models have been generalized to Randall-Sundrum brane
world models with warped extra dimensions, such that the effective Planck
scale again reduce to a $TeV$ scale \cite{rs12, rs14}. It has been argued
that such models motivated by string theory can be detected by analyzing the
QNMs obtained from gravitational antennas \cite{r1, r2, r3, r4}. However, as
the effective Planck scale is brought down in these models, it is important
to consider a UV completion of the effective theory in four dimensions for
analyzing such black holes. An important UV completion of the effective
field theory, obtained as a low energy limit of string theory is the
D-branes action, which can be described by a Born-Infeld (BI) action coupled
to dilaton field \cite{1985,Metsaev, Lovelace,1989}. So, in this paper, we
will analyze the QNMs from a black hole in such a theory with a BI action
coupled to a dilaton field.

It may be noted that another interesting UV completion of general
relativity, motivated by a formalism developed in condensed matter physics,
is the Horava-Lifshitz gravity \cite{HoravaPRD,HoravaPRL}. The space and
time scale with different Lifshitz scaling in the Horava-Lifshitz gravity,
such that it reduces to general relativity in the IR limit, but differs from
general relativity in the UV limit. In fact, motivated by Horava-Lifshitz
gravity, the consequences of such different Lifshitz scaling have been
studied for various geometrical structures that occur in the string theory.
Such a Lifshitz scale has been studied for both type IIA string theory \cite%
{A} and type IIB string theory \cite{B}. The behavior of dilaton black
branes \cite{1105.6335,1506.01784} and dilaton black holes \cite{d, d1} with
Lifshitz scaling has also been discussed. Furthermore, such black solutions
in a BI non-linear action with Lifshitz scaling have also been studied \cite%
{nl12}. The brane world models with warped extra dimensions and Lifshitz
scaling have also been constructed \cite{et, et12}.

It may be noted that even though the charged scalar perturbations are
important \cite{0207028,1405.4931,1406.0019,1002.2679,1010.2806,1111.6729},
mostly QNMs in Lifshitz theories have been studied for neutral scalar
perturbations \cite%
{1201.4521,0905.1136,1205.0582,1204.2116,1503.07457,1510.06012,1510.04605,1312.0826}%
. However, QNMs for a charged dilaton-Lifshitz solutions in the four
dimensions have also been studied \cite{MKZ}. As the QNMs can be used to
test models with large extra dimensions with Lifshitz scaling, it is
important to study the QNMs for higher dimensional analogs of such a black
hole. Furthermore, as the effective Planck scale is brought down in many
models with large extra dimensions, it is important to discuss such results
using a UV completion of such a theory. Thus, we will analyze the QNMs from
a higher dimensional dilaton-Lifshitz black hole solutions coupled to a
non-linear Born-Infeld action.

We organize this paper as follows: In section \ref{sec2}, we will review the
dilaton-Lifshitz black hole solutions in the presence of BI nonlinear
electrodynamics. In section \ref{sec3}, the wave equations of charged scalar
perturbations around our black holes will be presented. In latter section,
we will analyze the QNMs analytically. Then, we will use the asymptotic
iteration method to analyze QNMs numerically in section \ref{sec4}. Finally,
in the last section, we will summarize our results and discuss about them.

\section{Lifshitz Solutions\label{sec2}}

In this section, we will analyze a higher dimensional black hole in a UV
complete theory. Thus, we will consider a dilaton-Lifshitz black hole
coupled to non-linear BI action. The metric for such a $(n+1)$-dimensional
Lifshitz black holes can be expressed as \cite{1105.6335, 0905.1136} 
\begin{equation}
\mathrm{d}s^{2}=-\frac{r^{2z}f(r)}{l^{2z}}\mathrm{d}t^{2}+\frac{l^{2}}{%
r^{2}f(r)}\mathrm{d}r^{2}+r^{2}d\mathbf{\Omega }_{n-1}^{2},  \label{metric}
\end{equation}%
where $d\mathbf{\Omega }_{n-1}^{2}$ is a ($n-1$)-dimensional hypersurface
with constant curvature $(n-1)(n-2)$, such that 
\begin{equation*}
d\mathbf{\Omega }_{n-1}^{2}=d\theta _{1}^{2}+\sum\limits_{i=2}^{n-1}d\theta
_{i}^{2}\prod\limits_{j=1}^{i-1}\sin ^{2}\theta _{j}.
\end{equation*}%
Here $\omega _{n-1}$ is the volume and $z(\geq 1)$ is dynamical critical
exponent. Now it can be observed that in the limit, $r\rightarrow \infty $,
the metric given by Eq. (\ref{metric}) will asymptotically reduce to the
Lifshitz metric given by 
\begin{equation}
ds^{2}=-\frac{r^{2z}dt^{2}}{l^{2z}}+{\frac{l^{2}dr^{2}}{r^{2}}}+r^{2}d%
\mathbf{\Omega }_{n-1}^{2}.  \label{lifmet}
\end{equation}%
In this paper, we will analyze the coupling of a Lifshitz spacetime to a
non-linear BI action. So, first we observe that in the absence of dilaton
field, the function $L(F)$ for BI Lagrangian is given by \cite{BI7} 
\begin{equation}
L(F)=\frac{4}{\beta ^{2}}\left( 1-\sqrt{1+\frac{F\beta ^{2}}{2}}\right) ,
\end{equation}%
where $\beta $ is the BI non-linearity parameter, $F=F_{\mu \nu }F^{\mu \nu
} $ is the the Maxwell invariant, as $F_{\mu \nu }=\partial _{\mu }A_{\nu
}-\partial _{\nu }A_{\mu }$ with $A_{\mu }$ being the $U(1)$ abelian gauge
field. The dilaton field can couple to this $U(1)$ abelian gauge field. So,
in presence of a dilaton field, the Lagrangian for the BI coupled to a
dilaton scalar field $\Phi $ can be written as \cite{0801.4112,0709.3619}%
\begin{equation}
L(F,\Phi )=\frac{4}{\beta ^{2}}e^{4\lambda \phi /(n-1)}\left( 1-\sqrt{1+%
\frac{\beta ^{2}F}{2}e^{-8\lambda \Phi /(n-1)}}\right) ,
\end{equation}%
where $\lambda $ is a constant. In Einstein frame, the Lagrangian density of
Einstein-dilaton model with string theory corrections \cite{Polch}, contains
two $U(1)$ gauge field's \cite{1105.6335}. So, the Lagrangian for this
system can be written as 
\begin{equation}
16\pi \mathcal{L}=\mathcal{R}-\frac{4}{n-1}(\nabla \Phi )^{2}-2\Lambda
-\sum\limits_{i=1}^{2}e^{-4\Phi \lambda _{i}/(n-1)}H_{i}+L(F,\Phi ),
\label{Lag}
\end{equation}%
where $\mathcal{R}$ is the Ricci scalar and $\Lambda ,\lambda _{i}$ are
constants, in this Lagrangian. In the Lagrangian (\ref{Lag}), $H_{i}=\left(
H_{i}\right) _{\mu \nu }\left( H_{i}\right) ^{\mu \nu }$, where $\left(
H_{i}\right) _{\mu \nu }=2\partial _{\lbrack \mu }\left( B_{i}\right) _{\nu
]}$, and $\left( B_{i}\right) _{\mu }$ is a gauge field. In the limit, $%
\beta \rightarrow 0$, this Lagrangian $\mathcal{L}$ reduces to the
Einstein-dilaton-Maxwell Lagrangian (at the leading order) \cite%
{1105.6335,1506.01784}%
\begin{equation}
\lim_{\beta \rightarrow 0}16\pi \mathcal{L}=\cdots -e^{-4\lambda \Phi
/(n-1)}F+\beta ^{2}\frac{e^{-12\lambda \Phi /(n-1)}F^{2}}{8}+O\left( \beta
^{4}\right) .
\end{equation}%
By varying the action $S=\int_{\mathcal{M}}d^{n+1}x\sqrt{-g}\mathcal{L}$
with respect to the metric $g_{\mu \nu }$, the dilaton field $\Phi $, and $%
U(1)$ abelian gauge field $A_{\mu }$, and $\left( B_{i}\right) _{\mu }$, the
following field equations are obtained \cite{1610.06352}%
\begin{gather}
\mathcal{R}_{\mu \nu }-\frac{g_{\mu \nu }}{n-1}\left( 2\Lambda
+2L_{F}F-L(F,\Phi )-\sum\limits_{i=1}^{2}e^{-4\Phi \lambda
_{i}/(n-1)}H_{i}\right)  \notag \\
-\frac{4}{n-1}\partial _{\mu }\Phi \partial _{\nu }\Phi +2L_{F}F_{\mu
\lambda }F_{\nu }^{\text{ \ }\lambda }  \notag \\
-2\sum\limits_{i=1}^{2}e^{-4\lambda _{i}\Phi /(n-1)}\left( H_{i}\right)
_{\mu \lambda }\left( H_{i}\right) _{\nu }^{\text{ \ }\lambda }=0,
\label{FE1} \\
\nabla ^{2}\Phi +\frac{n-1}{8}L_{\Phi }+\sum\limits_{i=1}^{2}\frac{{\lambda }%
_{i}}{2}e^{-{4{\lambda }_{i}\Phi }/({n-1})}H_{i}=0,  \label{FE2} \\
\triangledown _{\mu }\left( L_{F}F^{\mu \nu }\right) =0,  \label{FE3} \\
\triangledown _{\mu }\left( e^{-{4\lambda }_{i}{\Phi }/({n-1})}\left(
H_{i}\right) ^{\mu \nu }\right) =0,  \label{FE4}
\end{gather}%
where we use the convention $X_{Y}=\partial X/\partial Y$. With metric (\ref%
{metric}), the Eqs. (\ref{FE3}) and (\ref{FE4}) can be solved for the $U(1)$
abelian gauge field, and thus we obtain 
\begin{eqnarray}
F_{rt} &=&\frac{qe^{4\lambda \Phi /(n-1)}r^{z-n}}{\beta \Gamma },
\label{Frt} \\
\left( H_{i}\right) _{rt} &=&q_{i}r^{z-n}e^{4\lambda _{i}\Phi /(n-1)},
\label{Hrt}
\end{eqnarray}%
where $\Gamma =\sqrt{1+q^{2}l^{2z-2}\beta ^{2}/(r^{2n-2})}$. By subtracting (%
$tt$), and ($rr$) components of Eq. (\ref{FE1}), $\Phi (r)$ can be obtained,
which is given by 
\begin{equation}
\Phi (r)=\frac{(n-1)\sqrt{z-1}}{2}\ln \left( \frac{r}{b}\right) ,
\label{Phi}
\end{equation}%
where $b$ is a constant. By substituting Eqs. (\ref{Frt}), (\ref{Hrt}), and (%
\ref{Phi}) in field equations (\ref{FE1}), and (\ref{FE2}), the $f(r)$
function for this system is obtained, and this given by 
\begin{widetext}
\begin{equation}
f(r)=\left\{ 
\begin{array}{ll}
1-\frac{m}{r^{n+z-1}}+\frac{k(n-2)^{2}l^{2}}{(n+z-3)^{2}r^{2}}+\frac{4l^{2}{b%
}^{2z-2}}{\beta ^{2}{r}^{2z-2}(n-1)(n-z+1)}-\frac{4l^{2}b^{2z-2}}{\beta
^{2}(n-1)r^{n+z-1}}\int \Gamma r^{n-z}{dr,} & \text{for }z\neq n+1, \\ 
&  \\ 
1-\frac{m}{r^{2n}}+\frac{k(n-2)^{2}l^{2}}{4(n-1)^{2}r^{2}}-\frac{4b^{2n}l^{2}%
}{\beta ^{2}(n-1)^{2}r^{2n}}\left[ 1-\Gamma +\ln \left( \frac{1+\Gamma }{2}%
\right) \right] , & \text{for }z=n+1.%
\end{array}%
\right.   \label{f4}
\end{equation}%
\end{widetext}Here $m$ is a constant, and it is related to the total mass of
black brane. Now as this system has to satisfy the field equations, we can
write 
\begin{gather}
\lambda =-\sqrt{z-1},\text{ \ \ \ \ }\lambda _{1}=\frac{n-1}{\sqrt{z-1}},%
\text{ \ \ \ \ }\lambda _{2}=\frac{n-2}{\sqrt{z-1}},  \notag \\
q_{1}^{2}=\frac{-\Lambda \left( z-1\right) b^{2(n-1)}}{\left( z+n-2\right)
l^{2(z-1)}},  \notag \\
q_{2}^{2}=\frac{(n-1)(n-2)(z-1)b^{2(n-2)}}{2(z+n-3)l^{2(z-1)}},  \notag \\
\Lambda =-\frac{(n+z-1)(n+z-2)}{2l^{2}}.  \label{constants}
\end{gather}%
Integrating the last term of $f(r)$ for $z\neq n+1$, we obtain 
\begin{align}
f(r)& =1-\frac{m}{r^{n+z-1}}+\frac{l^{2}(n-2)^{2}}{r^{2}(n+z-3)^{2}}  \notag
\\
& +\frac{4l^{2}b^{2z-2}r^{2-2z}\left( 1-\Gamma \right) }{\beta
^{2}(n-1)(n-z+1)}+\frac{4q^{2}b^{2z-2}l^{2z}r^{-2(n+z-2)}\Gamma }{%
(n-z+1)(n+z-3)}  \notag \\
& \times \mathbf{F}\left( 1,\frac{2n+z-4}{2n-2},\frac{3n+z-5}{2(n-1)}%
,1-\Gamma ^{2}\right) .
\end{align}%
We can observe from this solution, that in the limit $r\rightarrow \infty $, 
$f(r)$ satisfies $f(r)\rightarrow 1$ (note that $\mathbf{F}\left(
a,b,c,0\right) =1$). It may be noted that for $z=n+1$, a Schwartzshild-like
black hole does not exist in this system, because as $r\rightarrow 0$, $f(r)$
will go to positive infinity $f(r)\rightarrow +\infty $. Nevertheless, for $%
z\neq n+1$, the Schwartzshild-like black hole do exists in this system
(apart from the extreme and non-extreme black holes and naked
singularities). From here on, we fix $l$ and $b$ to unity.

\section{Quasi-normal Modes\label{sec3}}

Now in this section, we will analyze the QNMs for the solution discussed in
the previous section. So, first we consider a minimally coupled charged
massive scalar field in such a spherically symmetric black hole background.
It is assumed that this field satisfies the Klein-Gordon equation 
\begin{equation}
D^{\nu }D_{\nu }\Psi =m_{s}^{2}\Psi ,  \label{ChSc}
\end{equation}%
where $D^{\nu }=\nabla ^{\nu }-iq_{s}A^{\nu }$ is the gauge covariant
derivative and $m_{s}$ is the mass of the scalar field $\Psi $. Now we
decompose $\Psi $ into the following standard form, 
\begin{equation}
\Psi (t,r,\text{angles})=e^{-i\omega t}R(r)Y(\text{angles}),  \label{scf}
\end{equation}%
where $Y($angles$)$ is the spherical harmonic function related to the
angular coordinates, and $\omega $ is the frequency. As $Y$ is the spherical
harmonic function, it satisfies the following equation 
\begin{equation}
\nabla _{\text{angles}}^{2}Y(\theta ,\phi )=-QY(\text{angles}),
\end{equation}%
where $Q$ is the constant of separation. Moreover, the differential equation
for the radial function in the $(n+1)$-dimensional Lifshitz-dilaton
background (\ref{metric}) is%
\begin{eqnarray}
fR^{\prime \prime } &+&\left( f^{\prime }+\frac{(d+z-1)f}{r}\right)
R^{\prime }  \notag \\
&+&\left( \frac{q_{s}A_{t}+\omega }{r^{z+1}}\right) ^{2}\frac{R}{f}-\left(
m_{s}^{2}+\frac{Q}{r^{2}}\right) \frac{R}{r^{2}}=0,  \label{fr}
\end{eqnarray}%
where $d=n+1$. So, using a tortoise coordinate $r_{\ast }$ as $dr_{\ast }/dr=%
\left[ r^{z+1}f(r)\right] ^{-1}$, and introducing a new radial function $%
R(r) $ as $R(r)=K(r)/r$, the Eq. (\ref{fr}) can be transformed into an
equation resembling the Schr\"{o}dinger equation,%
\begin{equation}
\frac{d^{2}K(r_{\ast })}{dr_{\ast }^{2}}+\left[ \left( \omega
+q_{s}A_{t}(r)\right) ^{2}-V(r)\right] K(r_{\ast })=0,
\end{equation}%
where the potential $V(r)$ is given by 
\begin{eqnarray}
V(r) &=&\left( \frac{d^{2}}{4}+\frac{dz}{2}+1-d-z\right) f^{2}r^{2z}  \notag
\\
&+&fr^{2z}\left( m_{s}^{2}+\frac{Q}{r^{2}}\right) -ff^{\prime
}r^{2z+1}\left( 1-\frac{d}{2}\right) .
\end{eqnarray}%
Here we first observe that the first and the second terms are positive (with 
$m_{s}^{2}>0$). Now as $d\geq 4$, so the term $(1-d/2)$ is negative. Thus,
by having positive $f(r)$ and $f^{\prime }(r)$ out of the horizon $r>r_{h}$,
we can obtain $V(r)>0$ out of horizon provided that $m_{s}^{2}>0$. Note that 
$m_{s}^{2}$ can be negative as long as it is above
Breitenlohner-Freedman(-like) bound. One may read this bound from Eq. (\ref%
{bfbound}) in next section.

To perform a general stability analysis for this system, we first transform
to Eddington-Finkelstein coordinates, such that $v=t+r_{\ast }$. In this new
coordinate system, the metric can be written as 
\begin{equation}
ds^{2}=-r^{2z}fdv^{2}+2r^{z-1}dvdr+r^{2}d\mathbf{\Omega }_{n-1}^{2}.
\end{equation}%
Using the ansatz%
\begin{equation}
\Psi =e^{-i\omega v}\frac{\psi (r)}{r^{d/2-1}}Y(\text{angles}),
\end{equation}%
the differential equation for radial function $\psi (r)$ can be written as 
\begin{equation}
\left( r^{d+z-3}f\psi ^{\prime }\right) ^{\prime }-2i\left( \omega
+q_{s}A_{t}\right) \psi ^{\prime }-V\psi =0.
\end{equation}%
Multiplying above equation by $\bar{\psi}$ (the complex conjugate of $\psi $%
) and performing the integration from $r_{+}$ to infinity, we obtain%
\begin{equation}
\int_{r_{+}}^{\infty }dr\left[ r^{d+z-3}f\left\vert \psi ^{\prime
}\right\vert ^{2}+2i\left( \omega +q_{s}A_{t}\right) \bar{\psi}\psi ^{\prime
}+V\left\vert \psi \right\vert ^{2}\right] =0,  \label{eq}
\end{equation}%
where we used the integration by part and applied the Dirichlet boundary
condition for $\psi $ i.e., $\psi \left( \infty \right) \rightarrow 0$. For (%
\ref{eq}) to be satisfied, both the real and imaginary parts of integration
should vanish. Using the imaginary part, we observe\footnote{%
Note that $Im[2i\bar{a}b]=\bar{a}b+a\bar{b}$.} 
\begin{eqnarray}
0 &=&\int_{r_{+}}^{\infty }dr\left[ \left( \omega +q_{s}A_{t}\right) \bar{%
\psi}\psi ^{\prime }+\left( \bar{\omega}+\bar{q}_{s}\bar{A}_{t}\right) \psi 
\bar{\psi}^{\prime }\right]  \notag \\
&=&2i\int_{r_{+}}^{\infty }dr\mathrm{Im}[\left( \omega +q_{s}A_{t}\right) ]%
\bar{\psi}\psi ^{\prime }  \notag \\
&&-\bar{q}_{s}\int_{r_{+}}^{\infty }dr\left\vert \psi \right\vert ^{2}\bar{A}%
_{t}^{\prime }-\bar{\omega}\left\vert \psi \left( r_{+}\right) \right\vert
^{2}.  \label{eq1}
\end{eqnarray}%
Here we have again used the integration by part and the Dirichlet boundary
condition for $\psi $. In the case of neutral scalar field ($q_{s}=0$), Eq. (%
\ref{eq1}) reduces to%
\begin{equation}
\int_{r_{+}}^{\infty }dr\bar{\psi}\psi ^{\prime }=\frac{\bar{\omega}%
\left\vert \psi \left( r_{+}\right) \right\vert ^{2}}{2i\mathrm{Im}[\omega ]}%
.  \label{eq2}
\end{equation}%
Now we can substitute $\int_{r_{+}}^{\infty }dr\bar{\psi}\psi ^{\prime }$
from Eq. (\ref{eq2}) in Eq. (\ref{eq}) and obtain%
\begin{equation}
\int_{r_{+}}^{\infty }dr\left[ r^{d+z-3}f\left\vert \psi ^{\prime
}\right\vert ^{2}+V\left\vert \psi \right\vert ^{2}\right] =-\frac{%
\left\vert \omega \right\vert ^{2}\left\vert \psi \left( r_{+}\right)
\right\vert ^{2}}{\mathrm{Im}[\omega ]}.
\end{equation}%
Now we observe from this equation that for the potential $V$ to be positive
outside the horizon, imaginary part of $\omega $ is negative. So, for a
neutral scalar field, the black hole is stable under scalar field
perturbation, if the potential is positive outside the horizon (i.e. as long
as $m_{s}^{2}>0$). In next section, we proceed more to study charged scalar
fields numerically. We investigate the behavior of potential for $%
m_{s}^{2}<0 $, as well.

\section{Numerical Analysis\label{sec4}}

\begin{figure*}[t]
\centering
\subfigure[~$d=4, z=2$]
{\label{fig1a}\includegraphics[width=.32\textwidth]{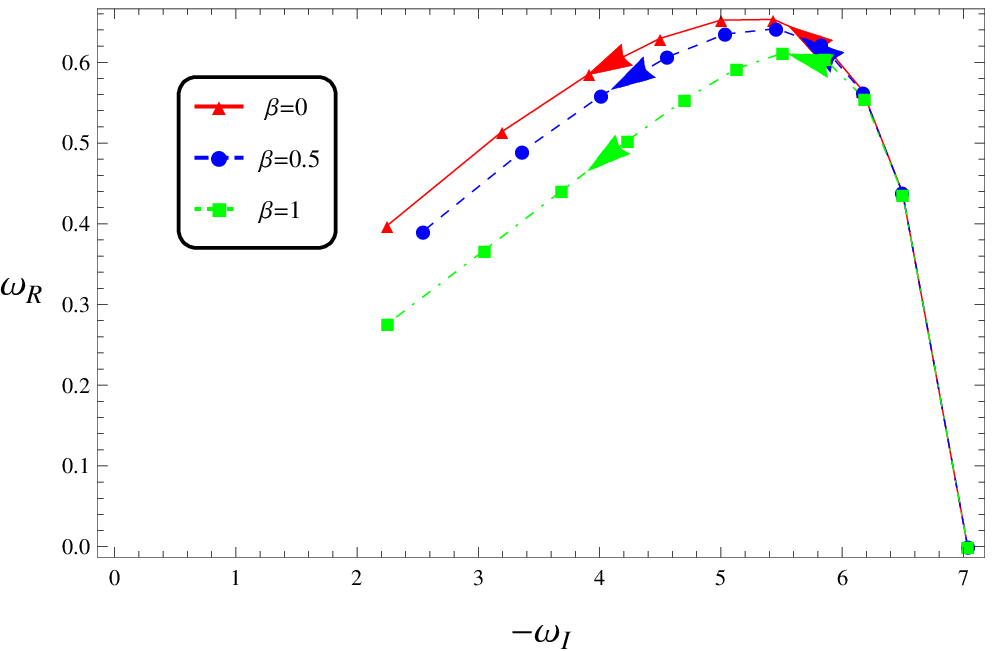} } %
\subfigure[~$d=4, z=3$]
{\label{fig1b}\includegraphics[width=.32\textwidth]{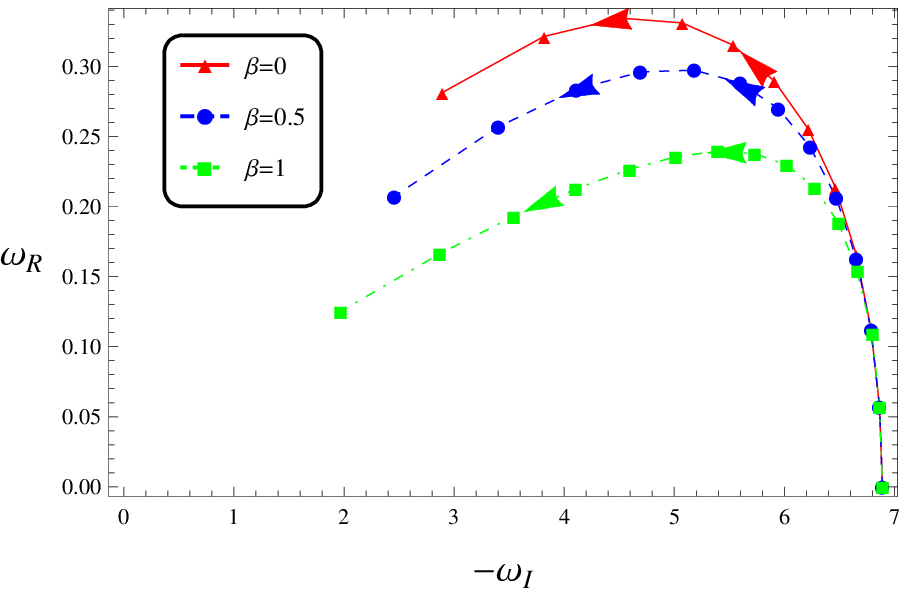} } %
\subfigure[~$d=5, z=2$]
{\label{fig1c}\includegraphics[width=.32\textwidth]{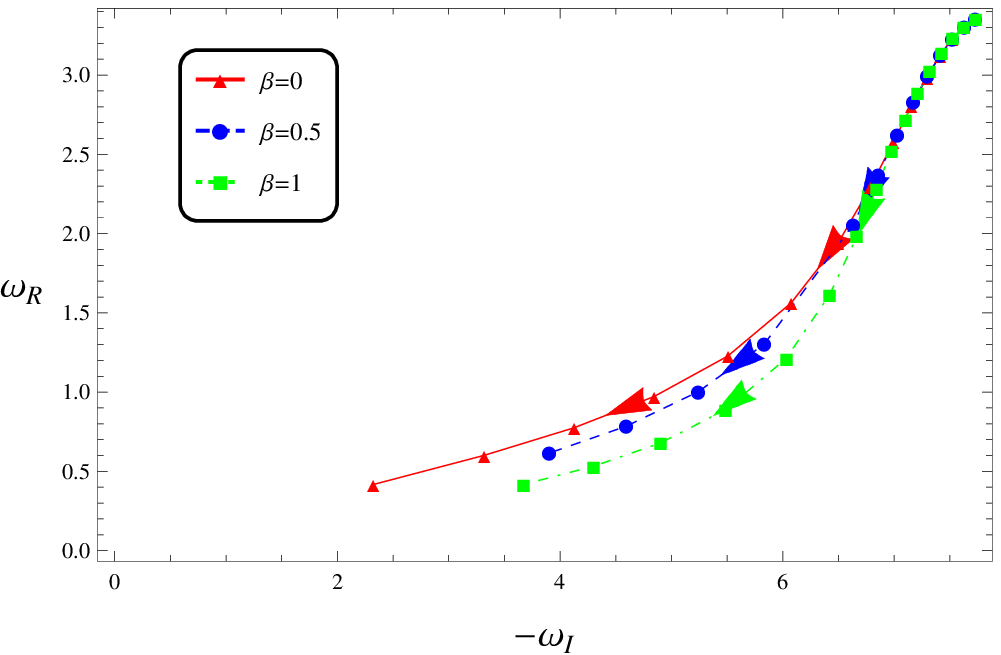} }
\caption{The behaviors of real and imaginary parts of QNFs in the case with $%
d\neq z$ and $m_{s}^{2}=1/4$ for different $d$, $z$, and $\protect\beta $.
The increase of the black holes charge $q$ from $0$ to extreme value are
shown with the arrows.}
\label{fig1}
\end{figure*}

\begin{figure*}[t]
\centering
\subfigure[~$d=4, z=4$]
{\label{fig2a}\includegraphics[width=.35\textwidth]{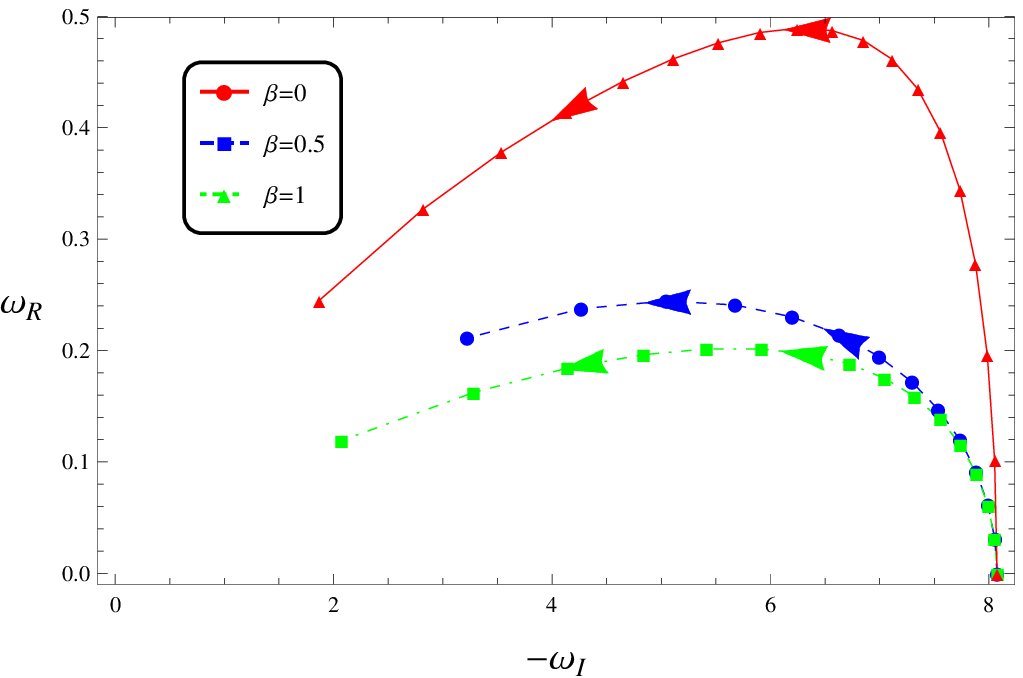} } \qquad %
\subfigure[~$d=5, z=5$]
{\label{fig2b}\includegraphics[width=.35\textwidth]{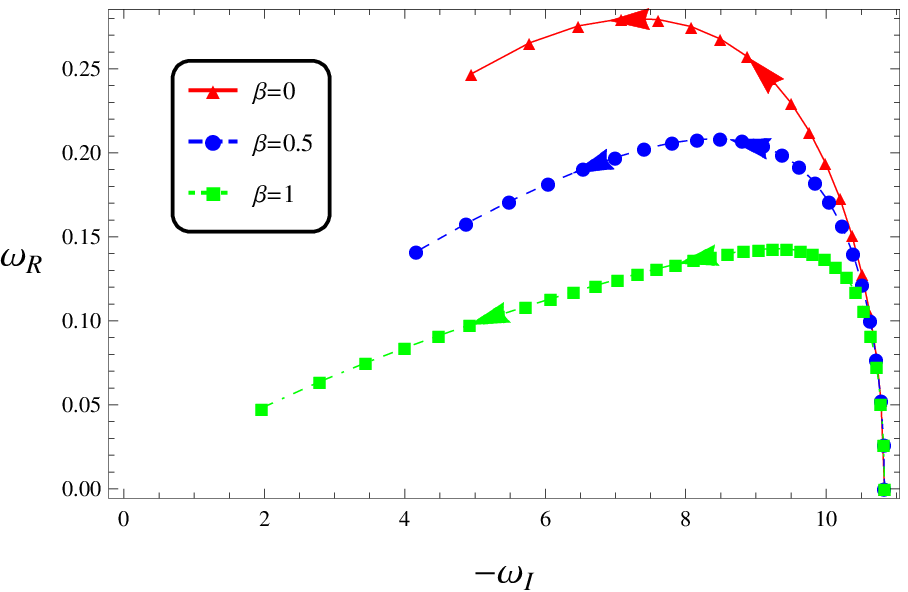} }
\caption{The behaviors of real and imaginary parts of QNFs in the case with $%
d=z$ and $m_{s}^{2}=1/4$ for different $\protect\beta $. The increase of the
black holes charge $q$ from $0$ to extreme value are shown with the arrows.}
\label{fig2}
\end{figure*}

In the previous section, we analyzed the stability of our black hole system,
when the potential was positive outside the horizon and scalar perturbation
was neutral. However, we cannot obtain such general results for charged
scalar fields as well as negative potential. To understand the behavior of
this system generally, we need to analyze it numerically. So, in this
section, we will analyze the behavior of QNMs numerically using the
asymptotic iteration method (AIM) \cite{0912.2740}. Now we first change the
variable $u=1-r_{h}/r$, and write Eq. (\ref{fr}) as 
\begin{eqnarray}
&&\frac{r_{h}^{z+1}f(u)}{(1-u)^{z-1}}R^{\prime \prime }(u)  \notag
\label{ru} \\
&&+\frac{r_{h}^{z+1}}{(1-u)^{z}}\left[ (d+z-3)f(u)+(1-u)f^{\prime }(u)\right]
R^{\prime }(u)  \notag \\
&&+\frac{r_{h}^{1-z}l^{2z+2}}{f(u)(1-u)^{1-z}}\left( q_{s}A_{t}(u)+\omega
\right) ^{2}R(u)  \notag \\
&&-\frac{r_{h}^{z+1}}{(1-u)^{z+1}}\left( m_{s}^{2}+\frac{Q}{r_{h}^{2}}%
(1-u)^{2}\right) R(u)=0.
\end{eqnarray}%
In order to propose an ansatz for (\ref{ru}), we are going to consider the
behavior of the function $R(u)$ at horizon $(u=0)$, and boundary $(u=1)$. At
horizon $(u=0)$, we have $f(0)\approx uf^{\prime }(0)$ and $A_{t}(0)=0$,
thus Eq. (\ref{ru}) can be written as 
\begin{equation}
R^{\prime \prime }(u)+\frac{R^{\prime }(u)}{u}+\frac{R(u)}{%
r_{h}^{2z}u^{2}f^{\prime 2}}\omega ^{2}=0.
\end{equation}%
The solution for this equation can be written as 
\begin{equation}
R(u\rightarrow 0)\sim C_{1}u^{-\xi }+C_{2}u^{\xi },\text{ \ \ \ \ }\xi =-%
\frac{i\omega }{r_{h}^{z}f^{\prime }(0)}.
\end{equation}%
Here we have imposed the ingoing boundary condition at the horizon $(u=0)$,
and so we have required $C_{2}$ to vanish.

At infinity, where $(u=1)$, Eq. (\ref{ru}), can be written as 
\begin{equation}
R^{\prime \prime }(u)+\frac{(z+d-3)R^{\prime }(u)}{1-u}-\frac{m_{s}^{2}R(u)}{%
(1-u)^{2}}=0.
\end{equation}%
The solution to this equation can be written as 
\begin{equation}
R(u\rightarrow 1)\sim D_{1}(1-u)^{\frac{1}{2}\left( z+2+\Pi \right)
}+D_{2}(1-u)^{\frac{1}{2}\left( z+2-\Pi \right) }.
\end{equation}%
where%
\begin{equation}
\Pi =\sqrt{(d+z-2)^{2}+4m_{s}^{2}}.  \label{bfbound}
\end{equation}%
In order to impose Dirichlet boundary condition $R(u\rightarrow
1)\rightarrow 0$, we can set $D_{2}=0$.

Using the above solutions at horizon and boundary, the general ansatz for
Eq. (\ref{ru}), can be written as 
\begin{equation}
R(u)=u^{-\xi }(1-u)^{\frac{1}{2}\left( z+2+\Pi \right) }\chi (u).
\label{Ruans}
\end{equation}%
Inserting Eq. (\ref{Ruans}) into (\ref{ru}), we obtain 
\begin{equation}
\chi ^{\prime \prime }=\lambda _{0}(u)\chi ^{\prime }+s_{0}(u)\chi ,
\label{AIM}
\end{equation}%
where the coefficient functions are given by 
\begin{equation}
\lambda _{0}=\frac{2i\omega }{r_{h}^{z}u+f^{\prime }(0)}-\frac{f^{\prime }(u)%
}{f(u)}+\frac{\Pi +1}{1-u}.
\end{equation}%
Furthermore, $s_{0}$ is given by%
\begin{eqnarray}
s_{0}(u) &=&\frac{r_{h}^{-2(z+1)}}{2(u-1)^{2}u^{2}f(u)^{2}f^{\prime 2}} 
\notag \\
&&\times \left( -2r_{h}^{2}u^{2}f^{\prime 2}(1-u)^{2z}(q_{s}A_{t}(u)+\omega
)^{2}\right.  \notag \\
&&+uf(u)f^{\prime }(0)r_{h}^{z}\left( 2uf^{\prime }(0)r_{h}^{z}\left(
m_{s}^{2}r_{h}^{2}+Q(u-1)^{2}\right) \right.  \notag \\
&&\left. +r_{h}^{2}(u-1)f^{\prime }(u)\right.  \notag \\
&&\left. \times \left( -uf^{\prime }(0)r_{h}^{z}\left( \Pi +d+z-2\right)
+2i(u-1)\omega \right) \right)  \notag \\
&&\left. +2r_{h}^{2}f(u)^{2}\left( i(u-1)\omega f^{\prime
}(0)r_{h}^{z}\left( u\Pi +1\right) \right. \right.  \notag \\
&&\left. \left. -m_{s}^{2}u^{2}f^{\prime 2}r_{h}^{2z}+(u-1)^{2}\omega
^{2}\right) \right) .
\end{eqnarray}%
Now using the improved AIM method, Eq. (\ref{AIM}) can be solved
numerically. We will derive the charged scalar perturbations, and study the
effects of the parameters of the model such as $d$, and $\beta $ on the real
and imaginary parts of quasi-normal frequencies. In the following, we will
set $Q=0$.

\subsection{Numerical results}

\begin{figure*}[t]
\centering
\subfigure[~$d= 4, z= 2$]
{\label{fig3a}\includegraphics[width=.35\textwidth]{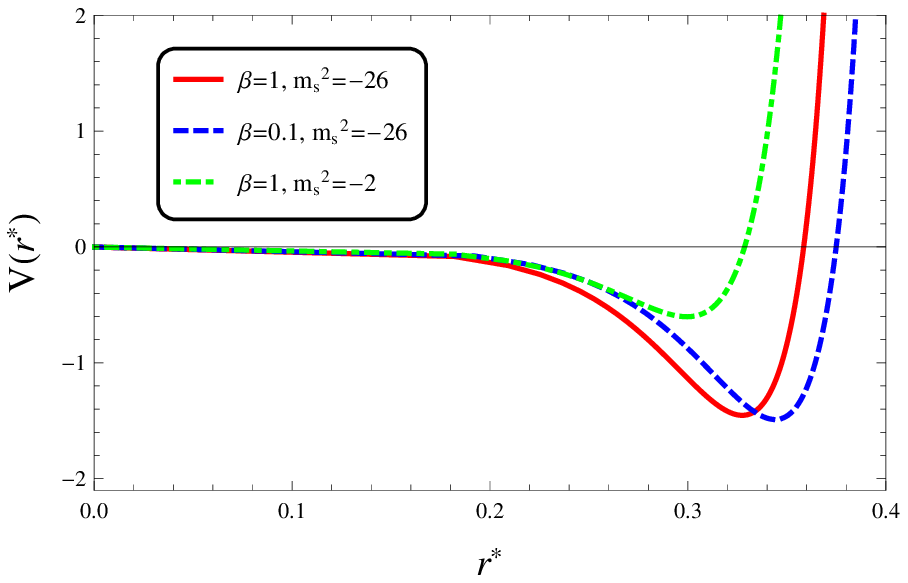} } \qquad %
\subfigure[~$d= 5, z= 3$]
{\label{fig3b}\includegraphics[width=.35\textwidth]{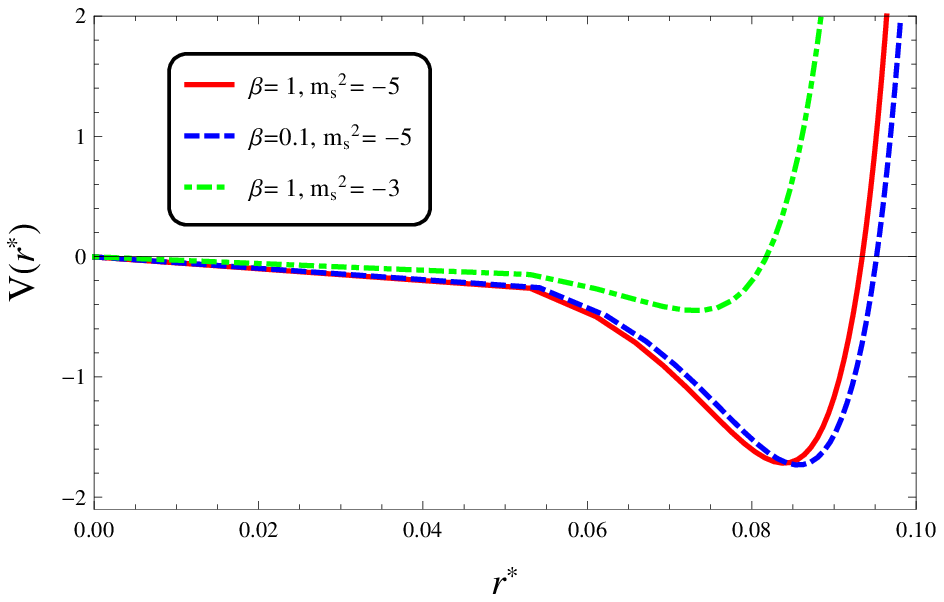} }
\caption{The behaviors of $V(r^{\ast })$ in the case with $d\neq z$ for
different $\protect\beta $ and $m_{s}^{2}$.}
\label{fig3}
\end{figure*}

\begin{figure*}[t]
\centering
\subfigure[~$d= 4, z= 4$]
{\label{fig4a}\includegraphics[width=.35\textwidth]{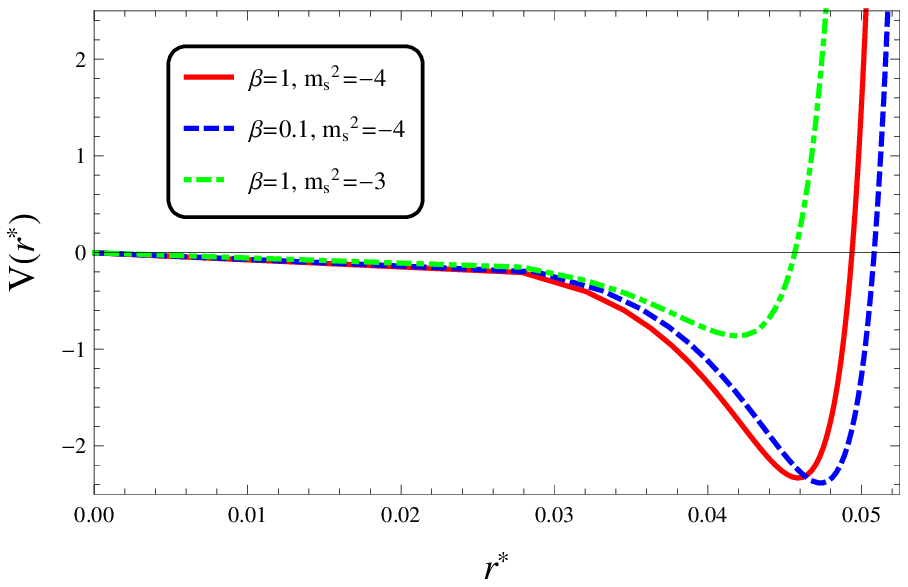} } \qquad %
\subfigure[~$d= 5, z= 5$]
{\label{fig4b}\includegraphics[width=.35\textwidth]{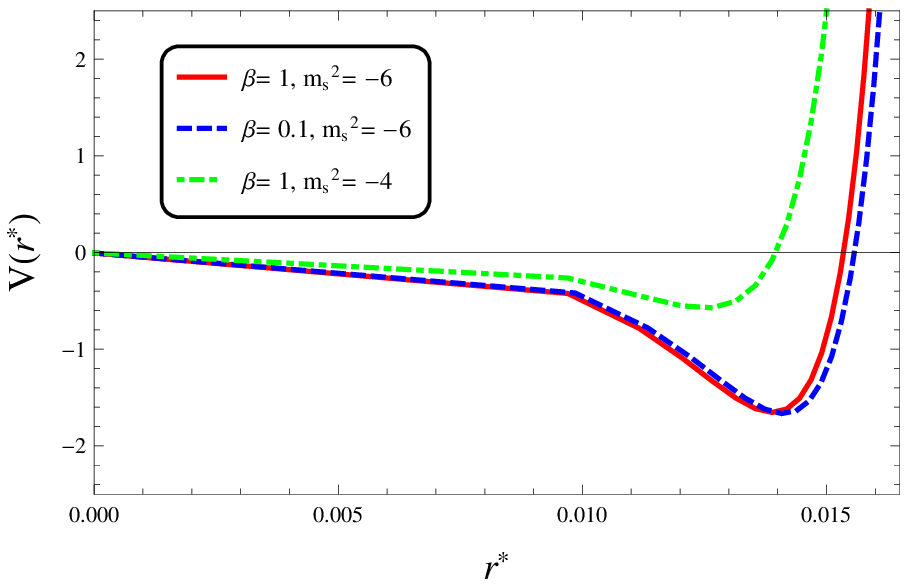} }
\caption{The behaviors of $V(r^{\ast })$ in the case with $d=z$ for
different $\protect\beta $ and $m_{s}^{2}$.}
\label{fig4}
\end{figure*}

In this section, we present the numerical results for the QNM of the charged
scalar perturbation around Lifshitz black hole and the effective potential $%
V(r^{\ast })$ as a function of tortoise coordinate $r^{\ast }$. We will
consider two cases, one for the case with $d\neq z$ (Figures \ref{fig1} and %
\ref{fig3}), and the other one with $d=z$ (Figures \ref{fig2} and \ref{fig4}%
).

The frequency of quasi-normal modes have both real and imaginary parts, $%
\omega =\omega _{R}+i\omega _{I}$. So, the factor $e^{-i\omega t}$ in scalar
field (Eq. (\ref{scf})) becomes $e^{-i\omega _{R}t}e^{\omega _{I}t}$.
Therefore, $\omega _{R}$ and $\omega _{I}$ determine the energy of scalar
perturbations and the stability of the system under dynamical
perturbationas, respectively. If $\omega _{I}>0$, the scalar field grows
with time and therefore the black hole is not stable under this
perturbation. If $\omega _{I}<0$, the black hole system is stable under
dynamical perturbation since the scalar field vanishes as time passes. For
larger values of $\left\vert \omega _{I}\right\vert $, the perturbation
remains longer outside the black hole untill it disappears. From holographic
point of view, it means that it takes more time for the dual system to go
back to equilibrium.

As one can see from Figures \ref{fig1} and \ref{fig2}, for both $d=z$ and $%
d\neq z$ cases, for fixed values of $\beta $, $\left\vert \omega
_{I}\right\vert $ decreases as $q$ increases (or equivalently temperature $T$
decreases). According to what explained above, it means that by increasing $%
q $, the complete decay of scalar perturbation outside the black hole takes
more time. Holographically, it shows that dual system needs more time to
return to equilibrium. As Figures \ref{fig1a} and \ref{fig1b} show, for $%
d\neq z$ case, by increasing $q$, $\omega _{R}$ increases up to a maximum
value first and then decreases. The value of maximum for a fixed $d$ and $z$%
, depends on nonlinear parameter $\beta $. For larger values of $\beta $,
the maximum value of $\omega _{R}$ is lower. It means that by increasing the
effects of nonlinearity, the maximum energy of perturbation decreases. For
specific values of $d$ and $z$, the influence of $\beta $ on real and
imaginary parts of quasi-normal frequencies disappears as $q$ tends to
smaller values (higher $T$). Moreover, for smaller values of $\beta $ but
fixed $d$ and $z$, the real part of quasi-normal frequencies is larger in
general as $q$ is further away from zero. The latter result shows that the
perturbation has more energy in linear electrodynamics case. Furthermore,
for larger values of $z$, the maximum energy of perturbations is lower
(Figures \ref{fig1a} and \ref{fig1b}). In $d=z$ case, we have the same
behaviors as $d\neq z$ case (Figures \ref{fig2a} and \ref{fig2b}). However,
as Figures \ref{fig1a} and \ref{fig1c} show, in the case of $d\neq z$, for
fixed $z$, by increasing $d$, the behavior of $\omega _{R}$ with respect to $%
q$ may change. For $d=5$ and $z=2$, $\omega _{R}$ shows a decreasing trend
as $q$ increases.

In the case with $z\neq d$ and $m_{s}^{2}<0$, the effective potential as a
function of $r^{\ast }$ are plotted in Figure \ref{fig3}. In Figure \ref%
{fig3a}, with $d=4$, and $z=2$, each potential has a well, whose height
depends on the value of $m_{s}^{2}$ and $\beta $. With the same $\beta $,
the one with lower $m_{s}^{2}$, has a higher hight of the well. The highest
hight of the well in this case is obtained when $\beta =0.1$ and $%
m_{s}^{2}=-26$. Moreover, the same behaviors is obtained by going to higher
dimension, such as $d=5$, as depicted in Figure \ref{fig3b}. This shows that
the particle described by the scalar field is traped in the potential of the
black hole in these cases. Another case is obtained by setting $d=z$, where
the dimension of spacetime equals the dynamical critical exponent $z$
(Figure \ref{fig4}). For $d=z$ cases, the same behavior is obtained as in
Figure \ref{fig3}, in which the potential has a well. In order to see the
effects of the existence of the extra dimension, in Figure \ref{fig4b}, we
consider the case $d=z=5$.

\section{Summary and Conclusion}

In this paper, we analyze the behavior of quasi-normal modes (QNMs) for a
higher dimensional black hole with Lifshitz scaling. This is important as
the QNMs can be used to test models with large extra dimensions with
Lifshitz scaling. In fact, as the effective Planck scale is lowered in such
models with large extra dimensions, we study these QNMs for a UV completion
action. This UV completion action is motivated by Born-Infeld action for a
D-brane action, which is a UV completion of the ordinary action for linear
actions of such systems. So, in this paper, the QNMs for higher dimensional
dilaton-Lifshitz black hole solutions coupled to a non-linear Born-Infeld
action have been studied. As it is important to study the charged
perturbations for such a black hole solution, such charged perturbations
were studied. In fact, we first analyzed general conditions for stability
analytically, for a positive potential. However, as it was not possible to
perform such an analysis for a charged perturbation as well as a negative
potential, we analyze this system for these cases numerically. This was done
using the asymptotic iteration method for quasi-normal modes. This is done
by analyzing this system for two cases, i.e., when $d\neq z$, and when $d=z$%
. For both cases, it is observed that the absolute value of imaginary parts
of quasi-normal frequencies decreases as charge of black hole increase to
extremal value. It means that it takes more time for QNMs to completely
decay outside the black hole. If one wants to interprete this in AdS/CFT
language, it shows that it needs more time for the system to go back to
equilibrium under the perturbation. Increasing the charge of black hole, the
real part of quasi-normal frequencies first increase up to a maximum and
then decreases. The value of this maximum is dependent on the nonlinearity
parameter $\beta $ and is higher for lower values of this. In the case of $%
d\neq z$, we have a different behavior for real part of quasi-normal
frequencies as the dimension of space-time increases. For $d=5\neq z$, we
observed that increase in value of black hole charge (equivalently decrease
in black hole temperature) cause the real part of quasi-normal frequencies
to decrease. To study the behavior of potential, we depicted the potential
in tortoise coordinate with $m_{s}^{2}<0$, for both $d\neq z$ and $d=z$
cases. We observed that potential is partly negative and has a well. For $d=4
$, each potential has a well whose hight depends on the $\beta $ and $%
m_{s}^{2}$. The same behavior is observed for $d=5$. It is observed that for
the second case, the high of well depends on $\beta $, and for a fixed $%
\beta $, the high depends on the value of $m_{s}^{2}$. Thus, the behavior of
this system depends both on $\beta $ and $m_{s}^{2}$ for a fixed dimension.

The QNMs have also become important in string theory, due to the development
of AdS/CFT correspondence \cite{adscft1, adscft2}. It has been observed that
the poles of the retarded Green function in a conformal field theory on the
boundary of an AdS space is related to the QNMs of a asymptotically AdS
black hole. Thus, QNMs of a higher dimensional AdS black hole can be used to
describe the behavior of strongly coupled quark-gluon plasmas \cite{qgp12,
qgp14}. As the AdS/CFT correspondence has been used to analyze the CFT dual
to a Lifshitz AdS spacetime \cite{adsl1, adsl2}, it would be interesting to
repeat the analysis done in this paper, for an asymptotically AdS black
hole. Then such QNMs can be used to analyze various properties of CFT dual
to such a black hole in the Lifshitz AdS spacetime. It would also be
interesting to analyze the effect that UV completion of the theory has on
the dual CFT. It may be noted that in this paper, we have used an abelian
Born-Infeld action, in which a $U(1)$ gauge field was coupled in a
non-linear action. The abelian Born-Infeld action has been generalized to a
non-abelian Born-Infeld action \cite{na12, na14}. Furthermore, black hole
solution in such a theory with non-abelian Born-Infeld have also been
constructed \cite{ab12, ab14}. It would be interesting to calculate the QNMs
for such black holes with a non-abelian Born-Infeld action. It would also be
interesting to analyze the QNMs for such black holes in an asymptotically
AdS spacetime, and then use the AdS/CFT correspondence to analyze the CFT
dual to such a system.

\begin{acknowledgments}
S. Sedigheh Hashemi would like to thank Shanghai Jiao Tong university for their warm hospitality during her visit,
and part of this project was done in Shahid Behehshti University.
This work has been supported financially by Research Institute for Astronomy
\& Astrophysics of Maragha (RIAAM) under research project No. 1/5237-53.
\end{acknowledgments}

\end{document}